\documentclass[sigconf,nonacm]{acmart}

\setcopyright{none}
\usepackage{amsmath,mathtools}
\usepackage{booktabs}
\usepackage{array}
\usepackage{graphicx}

\newcommand{\model}{\textsc{Frac-BKT}}
\newcommand{\fped}{F_{\mathrm{ped}}}
\newcommand{\mastery}{M}
\newcommand{\loss}{\mathcal{L}}

\begin{document}

\title{\model{} Under Mechanism Scrutiny: Response Assimilation and a Failed Fractional-Memory Test}

\author{Nizam Kadir}
\affiliation{%
  \institution{Singapore University of Technology and Design}
  \city{Singapore}
  \country{Singapore}}
\email{nizam_kadir@mymail.sutd.edu.sg}

\begin{abstract}
Long-horizon knowledge tracing requires more than an evocative memory mechanism: the
implementation must assimilate observed responses non-anticipatorily, and the mechanism must
survive matched controls. We subject \model{}, a BKT-style model with an
elapsed-time-scaled truncated fractional history, to both tests. The repaired model
predicts response $n$ from a prior mastery state and uses response $n$ only to update the
posterior state for later events. A five-seed evaluation uses 676,848 ASSISTments 2017
interactions from 1,694 learners and 19,249 learner--skill trajectories, with
chronological 80/10/10 suffix splits. The response-aware model reaches AUC .5860 overall,
.5638 after gaps of at least 30 minutes, and .5668 after at least 14 days. Relative to
the earlier non-response implementation, the gains are .0473, .0316, and .0277 AUC,
respectively. However, the fractional mechanism gate fails: the full model is not better
than the matched $\alpha=1$ control and is worse than the matched $\tau=0$ control by
.0090 AUC on both long-gap slices. Student-cluster wild-bootstrap intervals exclude zero,
both Holm-adjusted $p$-values are .00006, and all five per-seed differences favor $\tau=0$.
Compact AKT-, SAKT-, and DAS3H-style references also remain stronger on
long gaps. The contribution is therefore a response-assimilation repair and a falsifiable audit
showing that response assimilation helps while the evaluated fractional-history term
does not---not a state-of-the-art or causal-retention claim.
\end{abstract}



\maketitle

\section{Introduction}

An educational model can be accurate on the next click and still be uninformative about
what happens after the next week. This distinction is increasingly consequential as
AI-mediated tutoring makes immediate assistance cheap and ubiquitous. A learner may
receive productive scaffolding, or may obtain an answer without the retrieval and error
diagnosis needed for durable learning. Immediate correctness alone cannot distinguish
these cases. We call the resulting measurement risk---proximal success without verified
durable competence---the \emph{mastery-gain problem}. The phrase names an application
motivation, not an empirical finding of this paper: ASSISTments 2017 contains legacy
tutor hints rather than generative-AI interactions, and it does not contain randomized
delayed post-tests.

Knowledge tracing (KT) is a natural component of such an audit because it separates a
latent state from observed correctness. Bayesian Knowledge Tracing (BKT) is compact and
interpretable, but its state transition is Markovian \cite{corbett1994knowledge}. Deep
sequence models learn richer histories and often predict better
\cite{piech2015deep,pandey2019sakt,ghosh2020akt,choi2020saint}, yet their internal states
do not necessarily provide a small set of quantities that can be inspected as a
specified learning mechanism. Time-aware KT methods address forgetting and spacing
directly \cite{nagatani2019forgetting,choffin2019das3h,wang2021hawkeskt,im2023folibi},
but there remains room for models that make the assumed memory law explicit.

Fractional dynamics offer one such inductive bias. A Caputo derivative represents the
present rate of change as a power-law-weighted history of prior changes
\cite{podlubny1999fractional,metzler2000random,baleanu2016fractional}. Unlike a one-step Markov transition, the
resulting recurrence can distinguish trajectories with the same current state but
different recent state increments. Unlike unconstrained attention weights, the history
weights follow a declared analytic form. This does not make a fitted parameter a
biological measurement, but it does make the model's memory assumption inspectable.

We develop and stress-test an auditable fractional-state KT model, \model{}. The
claim-to-code audit materially narrows the claims relative to an accuracy-leaderboard
paper. The canonical experiment predicts later suffix events
from earlier events in the \emph{same} learner--skill trajectory; it does not evaluate
cold-start generalization to unseen learners. Its long-gap outcomes are next responses
after elapsed-time gaps; they are retention-oriented stress tests rather than direct
causal retention measurements. Its deep and time-aware comparators are compact,
controlled in-pipeline references, not faithful reproductions of the cited systems.

Our contributions are:
\begin{enumerate}
  \item We repair the knowledge-tracing semantics by predicting each response from a
  prior state and applying an exact BKT emission posterior only after that prediction.
  A non-anticipation test verifies that changing response $n$ cannot change prediction $n$ but
  does change later predictions.
  \item We give the exact implemented solver: a truncated uniform-lag L1 history with an
  event-specific elapsed-time scale \cite{stynes2022survey}. This makes clear both what the implementation
  computes and what it does \emph{not} claim to approximate exactly.
  \item We report a reproducible five-seed chronological-suffix evaluation with
  null-centered student-cluster wild bootstraps and pre-declared next-session and 14+ day slices.
  The results include negative comparisons and calibration failures, not only favorable
  AUC differences.
  \item We enforce matched $\alpha=1$ and $\tau=0$ mechanism gates under identical
  data, architecture, optimizer, seeds, and epochs. The completed controls falsify the
  fractional-history advantage: the no-history control is stronger on both pre-declared
  long-gap slices, with conditional learner-cluster intervals excluding zero.
\end{enumerate}

\section{Related Work}

\subsection{Knowledge Tracing and Time}

BKT models skill mastery with a hidden state and separates mastery from correctness
through guess and slip probabilities \cite{corbett1994knowledge,rabiner1989tutorial,badrinath2021pybkt}. Its standard
transition is memoryless after conditioning on the current state, a simplification that
conflicts with evidence that learning depends on elapsed time, repetition, and spacing
\cite{ebbinghaus1913memory,murre2015replication,cepeda2006distributed}. Deep Knowledge
Tracing and transformer variants relax this restriction by learning sequence
representations \cite{piech2015deep,pandey2019sakt,choi2020saint,ghosh2020akt,yin2023dtransformer}.
Benchmark work also shows that preprocessing, model-specified constraints, and implementation details can change model
rankings, making protocol disclosure part of the scientific result
\cite{xiong2016going,khajah2016deep,vaninwegen2019impact,liu2022pykt}.

Time-aware KT is already a substantial field. DAS3H uses recency and spacing windows;
HawkesKT models temporal excitation; forgetting-aware neural models inject elapsed-time
signals or recency biases \cite{choffin2019das3h,wang2021hawkeskt,nagatani2019forgetting,
im2023folibi}. Trainable spaced-repetition systems similarly connect elapsed time to
review decisions \cite{pashler2003temporal,settles2016trainable}. \model{} is not offered
as a replacement for this literature. Its narrower distinction is a latent-state
transition with an explicit fractional history and a compact exported diagnostic state.

\subsection{Assistance and Measurement}

Feedback and mastery-learning research emphasize durable capability rather than task
completion \cite{black1998assessment,bloom1968learning,shute2008feedback,
hattie2007power}. Intelligent tutors operationalize that goal through diagnosis and
scaffolding \cite{anderson1995cognitive,vanlehn2011relative,kulik2016effectiveness}.
Help seeking is beneficial only under constraints: assistance can support learning, but
overly direct help can enable shallow completion or gaming
\cite{aleven2003help,baker2004detecting,aleven2016help,inventado2018review}.
Generative AI intensifies this measurement problem
\cite{holmes2019artificial,kasneci2023chatgpt}. However, legacy hint counts are not an
identifying proxy for LLM use. We therefore use assistance telemetry only as a model
input and descriptive diagnostic, not as a causal treatment.

\subsection{Fractional Memory}

For $0<\alpha<1$, a Caputo derivative integrates prior rates of change with a power-law
kernel; $\alpha=1$ recovers an ordinary first derivative
\cite{podlubny1999fractional,metzler2000random,baleanu2016fractional}. Fractional operators have been used as
long-memory priors in physical systems and differentiable learning
\cite{wang2017fractional}. Their main computational cost is history: direct evaluation
is quadratic in sequence length. L1 discretization and finite-history truncation provide
a practical recurrence \cite{stynes2022survey}, but irregular timestamps require care. In particular, scaling a
uniform-lag L1 recurrence by the current gap is not the same as a mathematically derived
nonuniform-grid L1 method \cite{stynes2022survey}. We preserve that distinction throughout.

\section{Model}

\subsection{Observed Trajectories}

For learner $u$, skill $s$, and opportunity $n$, the log contains correctness
$y_{u,s,n}\in\{0,1\}$, hint count $h_{u,s,n}$, attempt count $a_{u,s,n}$, and timestamp
$t_{u,s,n}$. We model one trajectory per learner--skill pair. The latent state
$\mastery_{u,s,n}\in[0,1]$ is a model-defined mastery state, not an independently observed
test score.

We distinguish the prior state $\mastery_n^-$, used to predict response $n$, from the
posterior state $\mastery_n^+$, obtained only after observing that response. The
implemented pedagogical dose shifts hints and attempts by one event:
\begin{equation}
\label{eq:dose}
\fped{}_{n}=1+w^{(h)}_s\log(1+h_{n-1})
                 +w^{(a)}_s\log(1+a_{n-1}),
\end{equation}
with zero counts at the first event and nonnegative weights obtained by softplus. A
history context
\begin{equation}
\label{eq:gaming}
g_n=\frac{\sum_{k<n}h_k}{\max(\sum_{k<n}a_k,\epsilon)}
\end{equation}
summarizes cumulative assistance relative to attempts. We call this the
\emph{prior-assistance ratio}. The immutable canonical preprocessing artifact uses the
legacy field name \texttt{gaming\_index}; the repaired runner aliases that field before
it enters the model. We infer no gaming behavior from this ratio.

\subsection{Event-Conditioned Physics Parameters}

A skill embedding $e_s$ and the prior context $g_n$ enter a two-layer network:
\begin{equation}
(\tilde\alpha_n,\tilde\kappa_n,\tilde\lambda_n)
=f_\theta([e_s,g_n]).
\end{equation}
The constrained outputs are
\begin{align}
\alpha_n&=\operatorname{clip}(\sigma(\tilde\alpha_n),0.05,1),\\
\kappa_n&=\operatorname{softplus}(\tilde\kappa_n),\qquad
\lambda_n=\operatorname{softplus}(\tilde\lambda_n).
\end{align}
Thus $\alpha$, $\kappa$, and $\lambda$ are \emph{event-conditioned outputs}, not fixed
skill constants. Reported slice values are averages over events and seeds. Within the
specified model family, smaller $\alpha_n$ gives a heavier-tailed lag kernel, $\kappa_n$
scales conversion of dose into state gain, and $\lambda_n$ applies decay pressure. These
interpretations describe the recurrence; they do not establish cognitive constructs.

\subsection{Elapsed-Time-Scaled Fractional Recurrence}

The motivating continuous equation is
\begin{equation}
\label{eq:continuous}
{}_{0}^{C}D_t^{\alpha}\mastery(t)
=\kappa\fped{}(t)(1-\mastery(t))-\lambda\mastery(t).
\end{equation}
For event $n$, the implementation defines a transformed gap
\begin{equation}
\delta_n=\log\left(1+\frac{\max(t_n-t_{n-1},60\ \mathrm{s})}{60\ \mathrm{s}}\right)
\end{equation}
and an event scale
\begin{equation}
\label{eq:cscale}
C_n=\frac{\delta_n^{-\alpha_n}}{\Gamma(2-\alpha_n)}.
\end{equation}
It retains uniform-lag L1 weights
\begin{equation}
b_j(\alpha_n)=(j+1)^{1-\alpha_n}-j^{1-\alpha_n}
\end{equation}
for the most recent $\tau$ posterior-mastery increments:
\begin{equation}
\label{eq:history}
H_{n,\tau}=\sum_{j=1}^{\min(n-1,\tau)}
b_j(\alpha_n)(\mastery^+_{n-j}-\mastery^+_{n-j-1}).
\end{equation}
The bounded prior-state recurrence is
\begin{equation}
\label{eq:update}
\mastery_n^-=\Pi_{[0,1]}\left[
\frac{C_n\mastery^+_{n-1}-C_nH_{n,\tau}+\kappa_n\fped{}_n}
{C_n+\kappa_n\fped{}_n+\lambda_n}
\right].
\end{equation}
At the first event, the unavailable prior-state and history terms are omitted.

Equation~\eqref{eq:update} is an \emph{elapsed-time-scaled, uniform-lag L1-inspired
approximation} \cite{stynes2022survey}. It is not an exact nonuniform-grid discretization of
Equation~\eqref{eq:continuous}. The current gap changes $C_n$, while the history weights
still depend on opportunity lag $j$. We expose this hybrid explicitly because calling it
continuous time would overstate the numerical method.

With fixed $\tau$, the recurrence costs $O(T\tau)$ per trajectory and stores $O(T)$
states during training. Setting $\tau=0$ removes $H_{n,\tau}$. Setting $\alpha_n=1$
annuls all $b_j$ and yields the matched one-step architecture.

\subsection{Non-Anticipatory Emission and Posterior}

Skill-level guess and slip logits connect the prior state to the current response:
\begin{equation}
\label{eq:emission}
p_n=p(y_n=1)=\mastery_n^-(1-\operatorname{slip}_s)
 +(1-\mastery_n^-)\operatorname{guess}_s .
\end{equation}
After scoring $p_n$, the observed response updates the posterior by the exact BKT
emission rule:
\begin{equation}
\mastery_n^+=
\begin{cases}
\dfrac{\mastery_n^-(1-\operatorname{slip}_s)}{p_n}, & y_n=1,\\[5pt]
\dfrac{\mastery_n^-\operatorname{slip}_s}{1-p_n}, & y_n=0.
\end{cases}
\end{equation}
Thus $y_n$ can affect only events $n+1$ and later. An executable self-test changes one
response and verifies equality of all predictions through that event and a change in
later predictions. The run minimizes binary cross-entropy on training-suffix events:
\begin{equation}
\loss=-\sum_{n\in\Omega_{\mathrm{train}}}
[y_n\log p_n+(1-y_n)\log(1-p_n)].
\end{equation}
No additional regularizer is active. All models use final-epoch weights; there is no
validation-selected checkpoint. This design is simple and reproducible, but it also
limits claims about optimal model selection.

\section{Evaluation}

\subsection{Data and Split}

We use ASSISTments 2017 interaction logs
\cite{heffernan2014assistments,koedinger2013data}. The raw file contains 942,816 rows;
its SHA-256 digest is recorded in the artifact manifest.
After retaining valid binary-response rows and learner--skill trajectories with at least
15 opportunities, 676,848 interactions remain: 1,694 learners, 81 skills, and 19,249
trajectories.

Each trajectory is sorted by timestamp and stable row identifiers, then split
chronologically: the first 80\% of events are training, the next 10\% validation, and the
final 10\% test. This yields 533,756/67,155/75,937 rows. The first event in each
trajectory is not scored. Learners and skills therefore occur across partitions; the
generalization target is online prediction of later events within an existing
learner--skill trajectory, not prediction for held-out learners.

Two pre-declared test slices operationalize long-horizon stress tests. The
\emph{next-session} slice contains 6,386 responses after gaps of at least 30 minutes
(1,312 learners, 5,207 trajectories). The \emph{14+ day} slice contains 4,683 responses
(1,216 learners, 4,073 trajectories). These are delayed next-response outcomes. They are
not randomized delayed post-tests.

\subsection{Models and Training}

The response-aware suite uses fixed configurations, seeds 42, 101, 2026, 8888, and
12345, 15 epochs, batch size 64, Adam, and final-epoch evaluation. The full model uses
embedding dimension 16, learning rate 0.005, and $\tau=10$.

The controlled references include a differentiable BKT implementation and a separate
time-aware BKT whose transition depends on the transformed elapsed gap. Compact
TransformerEncoder implementations provide SAKT-style and AKT-style references; a
student/skill embedding model with five spacing windows provides a DAS3H-style reference;
and a compact recurrent temporal-excitation model provides a HawkesKT-style reference.
These names describe architectural inspiration. They are not official implementations
of the cited systems, and their scores must not be read as published benchmark rankings.

Two publication-critical controls use exactly the response-aware architecture:
\textsc{Response-$\alpha1$} fixes $\alpha_n=1$ with $\tau=10$, and
\textsc{Response-$\tau0$} retains dynamic $\alpha_n$ but removes the history tail. They
share response assimilation, preprocessing, embeddings, optimizer, learning rate,
batch size, epochs, seeds, and aggregation with the full model. The earlier
non-response \model{} is retained only as a forensic baseline: it exposes the benefit
of repairing response assimilation, not a matched test of fractional memory.

\subsection{Metrics and Inference}

We compute AUC, negative log-likelihood, Brier score, and 10-bin expected calibration
error \cite{naeini2015obtaining}. The main table reports the AUC of predictions averaged
across five seeds for each event, matching the unit used for paired inference. For each of
the six pre-specified full-versus-control comparisons, we form paired DeLong AUC influence
values, sum their differences within learner, apply an HC1 finite-cluster scale, and draw
99,999 null-centered Rademacher wild-bootstrap replicates. Two-sided Monte Carlo
$p$-values use $(e+1)/(B+1)$, where $e$ counts draws at least as extreme as the observed
studentized difference; step-down Holm adjustment covers all six comparisons. Inverted
null-error quantiles give the reported 95\% intervals. This inference remains conditional
on the fitted seed-averaged ensemble. We therefore also export all five paired per-seed
differences descriptively, without treating them as a population-level retraining test.

\section{Results}

\begin{table*}[t]
\caption{ASSISTments test AUC after averaging five predictions per event. The 30m+ and
14d+ columns are next responses after at least the stated gap. Reference rows are
compact in-pipeline implementations, not official benchmark reproductions.}
\label{tab:results}
\centering
\scriptsize
\begin{tabular*}{\textwidth}{@{\extracolsep{\fill}}lccc@{}}
\toprule
Model & Overall AUC & 30m+ AUC & 14d+ AUC \\
\midrule
AKT-style ref. & \textbf{.6275} & \textbf{.6313} & \textbf{.6308} \\
SAKT-style ref. & .5858 & .6126 & .6159 \\
DAS3H-style ref. & .5925 & .6113 & .6062 \\
HawkesKT-style ref. & .5901 & .5035 & .5081 \\
Differentiable BKT ref. & .5267 & .5201 & .5260 \\
Time-aware BKT ref. & .5242 & .5162 & .5220 \\
\midrule
Earlier non-response \model{} & .5387 & .5322 & .5391 \\
\textsc{Response-$\alpha1$} & .5855 & .5654 & .5695 \\
\textsc{Response-$\tau0$} & .5872 & \textbf{.5728} & \textbf{.5758} \\
\textbf{Full response-aware \model{}} & .5860 & .5638 & .5668 \\
\bottomrule
\end{tabular*}
\end{table*}

\subsection{Predictive Boundary}

Table~\ref{tab:results} separates model repair from ranking. Adding the ordered response
update raises the earlier \model{} by .0473 overall AUC (95\% CI
[.0434, .0512]), .0316 on 30m+ ([.0249, .0383]), and .0277 on 14d+
([.0202, .0347]). It also exceeds differentiable BKT by .0593 overall, .0437 on
30m+, and .0408 on 14d+, with all intervals excluding zero. These are large effects of
the repaired model as a whole; they do not isolate its fractional term.

The response-aware full model is not the strongest predictor. AKT-style exceeds it by
.0415 overall (.0414 prior to four-decimal rounding) and .0640 on 14d+. SAKT-style and DAS3H-style are higher on both long-gap
slices. The full model exceeds the compact HawkesKT-style reference after long gaps but
not overall. These controlled results rule out a state-of-the-art ranking claim.

\subsection{Mechanism Gate}

The mechanism gate fails. Relative to $\alpha=1$, the full model differs by
-.0016 AUC on 30m+ (95\% CI [-.0042, .0010], Holm $p=.44462$) and -.0027 on 14d+
([-.0057, .0002], Holm $p=.21960$); both intervals include zero and the per-seed signs
are mixed. Relative to $\tau=0$, it is worse by .0090 on 30m+
([-.0123, -.0057], Holm $p=.00006$) and .0090 on 14d+
([-.0126, -.0054], Holm $p=.00006$). All five paired seed differences favor $\tau=0$
on each delayed slice. Thus learned $\alpha$ is not shown to help, and the evaluated
history tail reduces long-gap discrimination under this protocol. The adjusted values
quantify learner-cluster uncertainty conditional on the fitted five-seed ensemble, not
uncertainty under retraining or alternative datasets.

\subsection{Visualizing the Falsification}

\begin{figure*}[t]
\centering
\includegraphics[width=\textwidth]{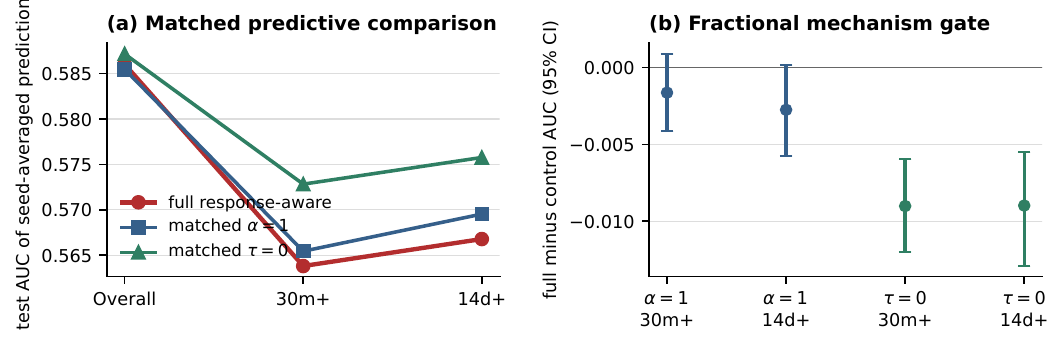}
\Description{Panel a compares test AUC for the full response-aware model and matched
alpha-one and no-history controls on overall, 30-minute, and 14-day slices. Panel b shows
full-minus-control AUC differences with 95 percent student-cluster wild-bootstrap intervals
for both long-gap slices.}
\caption{Matched response-aware evidence. (a) AUC of seed-averaged predictions.
(b) Full-minus-control differences with 95\% student-cluster wild-bootstrap intervals. The
history-free $\tau=0$ control is stronger on both long-gap slices, and both intervals
exclude zero.}
\label{fig:audit}
\end{figure*}

Figure~\ref{fig:audit} makes the publication gate explicit. The three response-aware
models are close overall, but the no-history control separates on delayed slices. This
conditional result uses predictions averaged over five seeds, the same non-anticipatory
posterior update and training protocol, and a paired student-cluster wild bootstrap.
Event-level $\alpha$, $\kappa$, and $\lambda$ remain available
as internal diagnostics, but parameter variation cannot be interpreted as evidence for
a mechanism that loses its matched ablation.

\section{Discussion}

\subsection{What the Evidence Supports}

The evaluation supports three claims. First, non-anticipatory response assimilation is essential:
the repaired model substantially outperforms the earlier implementation that did not
feed observed responses into future mastery. Second, the response-aware architecture
beats both BKT-family references on all three reported slices. Third, the matched
interventions identify a boundary that an uncontrolled comparison cannot: the tested
fractional-history tail does not improve long-horizon discrimination.

The evaluation does not support three stronger claims. It does not establish
state-of-the-art KT accuracy; compact attention and spacing references are stronger on
long gaps. It does not directly measure retained knowledge; the outcome is correctness
on the next logged response after a gap. It does not demonstrate that AI or hints cause
mastery inflation; no generative-AI treatment or randomized delayed post-test is
present.

This negative mechanism result is informative. A named parameter is not an
interpretation, and a continuous-time motivation does not show that its discrete
implementation helps. Mechanism-level interpretability requires matched interventions
that can falsify the proposed explanation.

\subsection{Implications for AI-Assisted Learning}

The mastery-gain problem remains a useful deployment motivation. A tutoring system that
optimizes only immediate response success can reward answer delivery rather than
learning. A retention-oriented KT layer could help schedule reassessment or identify
uncertainty after long gaps. However, the current model is not ready to generate learner
risk flags. Its calibration is weak, its parameters are conditional model outputs, and
its evidence comes from one legacy tutoring dataset. A responsible next study should
combine randomized assistance policies, delayed post-tests, direct AI interaction logs,
and pre-registered construct validation.

\subsection{Limitations and Threats to Validity}

\textbf{Construct validity.} Next-response-after-gap is only a proxy for retention.
Latent $\mastery_n$, $\alpha_n$, $\kappa_n$, and $\lambda_n$ are internal to the model.
Their labels do not make them independent cognitive measurements.

\textbf{Internal validity.} Hint and attempt counts are observational and may reflect
prior mastery, item difficulty, curriculum structure, or help-seeking policy. We make no
causal hint claim. The within-trajectory split conditions on earlier events from the same
learner and should not be confused with learner-held-out generalization.

\textbf{Numerical validity.} The hybrid recurrence scales uniform-lag L1 weights with the
current transformed gap. It is neither a constant-step experiment nor a derived
nonuniform L1 solver \cite{stynes2022survey}. A future numerical study should compare against a proper
nonuniform-grid method and test the one-minute gap floor.

\textbf{External validity.} Results use one dataset, only trajectories with at least 15
opportunities, and 71.8\% of raw rows. The in-pipeline reference models are not official
benchmark implementations. A second dataset and a library-controlled benchmark are
required before broad ranking claims.

\textbf{Selection and uncertainty.} Fixed configurations and final-epoch weights avoid
post-hoc test selection (rather than automated hyperparameter search \cite{akiba2019optuna}) but do not provide equal validation-search budgets. The paired
wild bootstrap operates on learner-cluster influence values after seed averaging. Per-seed
effects expose observed optimization sensitivity, but neither analysis captures uncertainty
from retraining on new samples, choosing another dataset, or changing model selection.

\section{Reproducibility and Ethics}

The internal audit archive records the raw-file hash, preprocessing and split manifest,
prediction and diagnostic files for all 15 response-aware model--seed jobs, 15
checkpoints, 225 unique history rows, metrics, and paired bootstrap outputs. The corrected
inference archive adds all 30 per-seed comparisons, six raw and Holm-adjusted tests, 8,444
privacy-preserving cluster-influence rows, and 599,994 retained null draws. Completed
jobs are written atomically and the runner is resumable. The non-anticipation self-test
and exact $\alpha=1$ and
$\tau=0$ tests run before training. Dataset, canonical-runner, and response-runner hashes
are stored in the completed metadata and independently rechecked.

The study analyzes de-identified educational interaction logs and does not intervene on
learners. Its main ethical risk is inferential: presenting an internal model parameter as
a learner trait could lead to inappropriate support or surveillance. We therefore avoid
learner-facing labels, causal language about hints, and deployment claims. Any future
use with generative-AI tutors should undergo institutional review, validate calibration
and subgroup performance, and keep pedagogical decisions contestable by learners and
educators.

\section{Conclusion}

The strongest result is not the one suggested by the model's name. Non-anticipatory response
assimilation produces a large and reproducible improvement over the earlier \model{}
implementation, but the full fractional-history model does not beat $\alpha=1$ and is
worse than $\tau=0$ after both 30 minutes and 14 days, with conditional learner-cluster
intervals excluding zero. Attention and
spacing references also remain stronger on long gaps. The broader lesson is
methodological: knowledge-tracing mechanisms should be executable, temporally ordered,
and subjected to matched interventions. An interpretable parameter earns scientific
meaning only when the mechanism it names survives that audit.

\begin{acks}
The author thanks Prof.\ Ricky Lay Kee Ang, Dr.\ Kee Chun-Yun, and Nguyen Gia Tuyen at the Singapore University of Technology and Design for motivating discussions on fractional dynamics and independent replication checks following the July 2026 canonical evaluation archived here.
\end{acks}

\bibliographystyle{ACM-Reference-Format}
\bibliography{references}

\clearpage
\appendix

\section{Idealized L1 Derivation}

For a uniform grid $t_n=n\Delta t$, the Caputo derivative is
\begin{equation}
{}_{0}^{C}D_t^\alpha\mastery(t_n)
=\frac{1}{\Gamma(1-\alpha)}
\int_0^{t_n}(t_n-\xi)^{-\alpha}\mastery'(\xi)\,d\xi .
\end{equation}
Approximating $\mastery'(\xi)$ as constant on each interval gives the standard L1 discretization \cite{stynes2022survey}:
\begin{align}
{}_{0}^{C}D_t^\alpha\mastery(t_n)
&\approx \frac{\Delta t^{-\alpha}}{\Gamma(2-\alpha)}
\bigg[(\mastery_n-\mastery_{n-1}) \nonumber\\
&\quad+\sum_{j=1}^{n-1}b_j(\alpha)
(\mastery_{n-j}-\mastery_{n-j-1})\bigg].
\end{align}
where $b_j(\alpha)=(j+1)^{1-\alpha}-j^{1-\alpha}$. Substitution into
Equation~\eqref{eq:continuous} and collection of terms in $\mastery_n$ yields the
idealized implicit update corresponding to Equation~\eqref{eq:update}. The canonical
implementation replaces the common $\Delta t$ scale by the event-specific transformed
gap in Equation~\eqref{eq:cscale} while retaining the uniform-lag weights. This last
replacement is an engineering approximation, not part of the uniform-grid derivation.

\section{Response-Aware Protocol Details}

The artifact manifest records the 942,816-row raw file and its SHA-256 digest. The
retained sample has 676,848 rows, 1,694 learners, 81 skills, and 19,249 trajectories.
The split is chronological 80/10/10 within trajectory and excludes each trajectory's
first event from scoring. Training uses seeds 42, 101, 2026, 8888, and 12345 for 15
epochs with batch size 64 and final-epoch evaluation. The response-aware models use
embedding dimension 16 and Adam with learning rate .005; the full and $\alpha=1$
models use $\tau=10$, while the no-history control uses $\tau=0$. The corrected paired
inference uses 99,999 null-centered Rademacher wild-bootstrap draws per comparison after
seed averaging, with learner-clustered DeLong influence values, finite-Monte-Carlo
correction, and step-down Holm adjustment. Note that while Table~\ref{tab:results} reports the AUC of the seed-averaged ensemble predictions (.5860 for full response-aware \model{}, .5855 for $\alpha=1$, and .6275 for AKT-style), the corresponding arithmetic means of the five individual per-seed AUCs from the same July 2026 evaluation runs are .5853 for full response-aware \model{}, .5845 for the matched $\alpha=1$ control, and .6210 for the AKT-style reference. All 15 prediction files,
15 diagnostic files, 15 checkpoints, 225 history rows, the split manifest, metrics,
bootstrap results, metadata, and hashes are retained in the internal audit archive.
In the artifact glossary, long-horizon denotes a next response after a pre-declared gap,
and named parameters denote event-conditioned recurrence outputs. Style references are
compact in-pipeline architectures; matched controls change only $\alpha=1$ or $\tau=0$
inside the same non-anticipatory response-aware implementation.
None of these terms denotes a causal effect of hints or generative AI.

\begin{table}[b]
\caption{Corrected matched-control inference. Differences are full response-aware
\model{} minus control; Holm adjustment covers all six rows. Seed counts are descriptive.}
\label{tab:inference}
\centering
\scriptsize
\setlength{\tabcolsep}{2.8pt}
\begin{tabular}{@{}llrrrr@{}}
\toprule
Control & Slice & $\Delta$AUC [95\% CI] & Raw $p$ & Holm $p$ & Seeds \\
\midrule
$\alpha=1$ & Overall & .00055 [-.00060, .00169] & .34486 & .44462 & 4/5 \\
$\alpha=1$ & 30m+ & -.00163 [-.00425, .00099] & .22231 & .44462 & 3/5 \\
$\alpha=1$ & 14d+ & -.00275 [-.00571, .00019] & .06820 & .21960 & 4/5 \\
$\tau=0$ & Overall & -.00114 [-.00230, .00002] & .05490 & .21960 & 2/5 \\
$\tau=0$ & 30m+ & -.00902 [-.01230, -.00572] & .00001 & .00006 & 5/5 \\
$\tau=0$ & 14d+ & -.00898 [-.01260, -.00538] & .00001 & .00006 & 5/5 \\
\bottomrule
\end{tabular}
\end{table}

\end{document}